\documentclass{ws-procs9x6}
\newcommand{\eqb}{\begin{equation}}
\newcommand{\eqe}{\end{equation}}
\newcommand{\dmb}{\begin{displaymath}}
\newcommand{\dme}{\end{displaymath}}
\newcommand{\pd}{\partial}

\newcommand{\eab}{\begin{eqnarray}}
\newcommand{\eae}{\end{eqnarray}}

\newcommand{\be}{\begin{equation}}
\newcommand{\ee}{\end{equation}}

\begin{document}

\title{The deconfining phase of SU(2) Yang-Mills thermodynamics}

\author{Ralf Hofmann$^*$}

\address{Universit\"at Heidelberg\\ 
Institut f\"ur Theoretische Physik\\ 
Philosophenweg 16\\ 
69120 Heidelberg, Germany\\
$^*$E-mail: r.hofmann@thphys.uni-heidelberg.de\\
http://www.thphys.uni-heidelberg.de/~hofmann/}

\begin{abstract}

We discuss nonperturbative SU(2) Yang-Mills thermodynamics in the deconfining phase. 
The maximal 
resolution of trivial-topology fluctuations 
is set by coarse-grained, interacting calorons and anticalorons: 
The effective loop expansion is very efficient. 
Postulating that SU(2)$_{\tiny\mbox{CMB}}\stackrel{\tiny\mbox{today}}=$U(1)$_Y$, a 
modification of thermalized, low-momentum photon propagation is predicted 
for temperatures a few times 2.7\,K. Phenomenological implications are: 
magnetic-field induced dichroism and birefringence at a temperature of 4.2\,K (PVLAS), stability of 
cold and dilute H1 clouds, and absence of low-$l$ 
correlations in the TT CMB power spectrum.    

\end{abstract}

\keywords{caloron, holonomy, coarse-graining, polarization tensor}

\bodymatter

\section{A nonperturbative ground state at high temperature}\label{sec1}

In spite of the innocent-looking title I will point out 
that the deconfining phase of an SU(2) 
Yang-Mills theory relies, in an essential way, on interspersed nonperturbative delicacies. 
As a consequence, unexpected results emerge in low-temperature, 
low-momentum photon propagation. 

Let me first sketch the phase diagram of an SU(2) (or SU(3)) 
Yang-Mills theory as derived in \cite{Hofmann2005}: There are a deconfining, 
high-temperature phase, a preconfining, thin intermediate phase, 
and a low-temperature confining phase. In the former two phases excitations 
are partially and exclusively massive gauge bosons, respectively. 
In contrast to perturbative screening temperature-dependent 
masses are induced by topological field configurations 
upon a spatial coarse-graining: Masses 
appear at tree-level by Higgs mechanisms in the associated effective theories. 
In the confining phase excitations are spin-1/2 fermions with equidistant mass 
spectrum set by the Yang-Mills scale\footnote{Except for a small range of 
temperatures above the critical temperature $T_{c}$ the 
pressure is positive in the deconfining phase and reaches the 
Stefan-Boltzmann limit in a power-like way. While the total pressure 
is negative in the preconfining phase it is precisely zero at $T=0$.}. 

Accurate results are obtained by (i) the consideration 
of BPS saturated, topological configurations (calorons and anticalorons) and 
(ii) a self-consistent spatial coarse-graining. An essential singularity in the 
weight of (anti)calorons at zero coupling forbids them 
in weak-coupling expansions. This is at the heart 
of the magnetic-sector instability 
encountered in perturbation theory \cite{Linde1980,Polyakov1975}. 

One writes a (unique) definition for the kernel ${\cal K}$ of a to-be-determined 
differential operator ${\cal D}$ in terms of the composite 
%**********
\eqb
\label{adjointcc}
\mbox{tr}\,\frac{\lambda^a}{2}\,F_{\mu\nu}\,\left((\tau,0)\right)\,
\left\{(\tau,0),(\tau,\vec{x})\right\}\,F_{\mu\nu}\,\left((\tau,\vec{x})\right)\,
\left\{(\tau,\vec{x}),(\tau,0)\right\}\
\eqe
%************
of fundamental field variables. 
In (\ref{adjointcc}) $F_{\mu\nu}$ is the field strength and 
$\left\{(\tau,0),(\tau,\vec{x})\right\}$ denotes a fundamental Wilson line\footnote{On the level of 
BPS saturated configurations, see below, no scale is available for a shift $0\to \vec{y}$. 
As a consequence, the definition in (\ref{adjointcc}) is no restriction of generality.}. 
${\cal K}$ contains the phase $\hat{\phi}^a$ of an emerging 
adjoint scalar field $\phi^a$. Due to an indefinite spatial coarse-graining 
$\hat{\phi}^a$ depends, in a periodic way, only on euclidean time $\tau$: 
No information on dimensional transmutation enters in $\hat{\phi}^a(\tau)\in {\cal K}$. 
Therefore, it suffices to evaluate the object in  (\ref{adjointcc}) on 
absolutely stable classical configurations and the Wilson lines are along straight lines\footnote{No scale determining a curvature 
of a spatial path is available.}. But only BPS saturated configurations 
are absolutely stable\footnote{All other solutions have higher euclidean action: A 
departure from classical trajectories takes place by 
their decay into BPS saturated plus 
topologically trivial configurations.}. In passing I mention that adjointly transforming 
{\sl local} composites vanish on BPS saturated configurations. For 
$Q=0$ BPS saturated configurations are pure gauges, $F_{\mu\nu}\equiv 0$, and (\ref{adjointcc}) 
vanishes identically. For 
$|Q|=1$ stable BPS saturated configurations are 
trivial-holonomy or Harrington-Shepard (HS) \cite{HS1977,Diakonov2004} (anti)calorons
\footnote{Each solution enters the definition (\ref{adjointcc}) separately, and the sum over $Q=\pm1$ is taken 
subsequently.}. 
The integration over the independent\footnote{The integral over global spatial or color rotations 
is contained in the spatial average because of the particular 
structure of the HS (anti)caloron. Nontrivial periodicity excludes the integration over 
time translations. The integration over space translations leaves ${\cal K}$ invariant because each shift is compensated for 
by an according parallel transport: This integration is already performed.} moduli of HS 
(anti)calorons (scale parameter $\rho$) must be subject to a flat measure since 
no scale exists which would set a `spectral slope' for this dimensionless quantity. Also, 
there is no a priori  cutoff for the spatial coarse-graining. It is easily checked by 
dimensional counting that both adding higher $n$-point functions of the field strength 
to (\ref{adjointcc}) and BPS saturated configurations with $|Q|>1$ are forbidden 
(dimensionful space and moduli integrations). Thus ${\cal K}$ is defined by integrating 
the $Q=\pm 1$ sum of the expression in (\ref{adjointcc}) 
with the weight $\int d^3x\int\rho$. 

In the radial ($r$) part of space integral a logarithmic divergence 
occurs for the magnetic-to-magnetic correlation of the field 
strength \cite{Hofmann2005,HerbstHofmann2004}. At the same time, the 
azimuthal angular integration yields zero. The former divergence 
can be regularized in a rotationally invariant way (dimensional regularization). This is not true for the 
latter zero: an apparent breaking of rotational symmetry is required for regularization. 
Namely, a defect (or surplus) angle needs to be defined with respect 
to a fixed direction in the azimuthal plane. Since distinct directions are connected by global gauge rotations 
no breaking of rotational symmetry is 
detected in a physical quantity. Thus the angular regularization is admissible. 

Performing the integrals, undetermined normalizations appear for each contribution (caloron or anticaloron). 
Moreover, there are undetermined global phase shifts $\tau\to \tau+\tau_{C,A}$. 
The convergence towards ${\cal K}=\left\{\hat{\phi}^a\left|D\hat{\phi}^a\equiv\left[\pd_\tau^2+\left(\frac{2\pi}{\beta}\right)^2\right]\hat{\phi}^a=0;\, 
\mbox{fixed ang. reg.}\right.\right\}$ is extremely fast. That is, with finite upper limits $\rho_u$ and $r_u$ in  
both the $\rho$- and the $r$-integration the $\tau$-dependence of the results resembles the limiting behavior 
($\rho_u=r_u=\infty$) within a small error already for 
$\rho_u$ and $r_u$ a few times $\beta\equiv 1/T$ \cite{Hofmann2005,HerbstHofmann2004}. 
This, however, makes the introduction 
of a finite cutoff $|\phi|^{-1}$ self-consistent: At fixed 
global gauge the infinite-volume coarse-graining, determining the $\tau$-dependence 
of $\hat{\phi}^a$, is saturated on a finite ball of radius $\sim |\phi|^{-1}$. 

How large is $|\phi|^{-1}$? Since a sufficiently large 
cutoff $|\phi|^{-1}$ saturates ${\cal K}$, since $D\hat{\phi}^a=0$ is a linear equation, 
and since $|\phi|$ is $\tau$-independent\footnote{Composed of coarse-grained, large quantum fluctuations 
$\Rightarrow$ no finite Matsubara frequencies.} we also have $D\phi=0$. Moreover, since a (finite) coarse-graining over 
noninteracting, BPS saturated configurations implies the BPS saturation of the field 
$\phi$ we need to find an appropriate square root of $D\phi=0$. Assuming the existence of a scale 
$\Lambda$, which together with $\beta$ determines the scale $|\phi|$, the right-hand side 
of the BPS equation must not depend on $\beta$ explicitly and must be analytic and linear in $\phi$. 
The only consistent option (up to global gauge rotations) is
$\pd_\tau\phi=\pm i\lambda_3\Lambda^3\phi^{-1}$ where 
$\phi^{-1}\equiv\frac{\phi}{|\phi|^2}$. Solutions are $\phi(\tau)=\sqrt{\frac{\Lambda^3\beta}{2\pi}}\,\lambda_1\,
\exp\left(\mp\frac{2\pi i}{\beta}\lambda_3(\tau-\tau_0)\right)$ where $\tau_0$ is a physically irrelevant 
integration variable (global gauge rotation). 
A critical temperature $2\pi T_c=13.867\,\Lambda$ exists, see \cite{Hofmann2005}. 
Thus, expressing the cutoff $|\phi|^{-1}=\sqrt{\frac{2\pi}{\Lambda^3\beta}}$ in units of $\beta$, yields 
$8.22$ at $T_c$; for $T>T_c$ this number grows as $(T/T_c)^{3/2}$. But for $\rho_u\sim r_u\ge 8.22$ the kernel 
${\cal K}$ is practically that of the infinite-volume limit, see also \cite{Hofmann2005,Herbst2005,HerbstHofmann2004}. 

Coarse-graining the $Q=0$ sector alone, leaves the Yang-Mills action form-invariant\footnote{ 
By all-order perturbative renormalizability interaction effects are absorbed into 
re-definitions of the parameters of the bare 
action \cite{'tHooftVeltman}.}. One can shown that the field $\phi$ 
is inert: Quantum fluctuations of 
resolution $<|\phi|$ do not deform $\phi$ making it a background for the 
coarse-grained $Q=0$-dynamics\cite{Hofmann2005}. The gauge-invariant extension of the kinetic term 
$\mbox{tr}\,\left(\pd_\tau\phi\right)^2$ in the (gauge-dependent) action for the field $\phi$ alone 
is $\pd_\tau\to D_\tau$ ($D_\tau$ the adjoint covariant derivative): A 
unique effective action emerges. The equations of motion for the 
$Q=0$ sector (subject to the coarse-grained $|Q|=1$-background) 
possess a pure-gauge solution $a_\mu^{bg}$: The ground-state energy density $\rho^{gs}$ and 
pressure $P^{gs}$ then are $\rho^{gs}=-P^{gs}=4\pi\Lambda^3\,T$\footnote{A negative ground-state 
pressure is expected microscopically due to the dominating dynamics of 
small-holonomy calorons leading to finite life-time cycles 
of magnetic dipoles (a magnetic monopoles attracts its antimonopole, 
the pair annihilates, and is recreated)\cite{Diakonov2004}.}.  

\section{Constraints on resolution in the effective theory}\label{sec2}

Two color directions acquire mass (adjoint Higgs mechanism). In unitary gauge, 
$\phi=\lambda_3\,|\phi|\,,\ \ a_\mu^{bg}=0$, one has $m_{1,2}\propto |\phi|$. Fixing the remaining U(1) by 
$\pd_i a^{a=3}_i=0$ a given mode's momentum is physical. To distinguish between quantum and thermal 
fluctuations we work in the real-time formalism when integrating out gauge-field 
fluctuations in the effective theory. 

Two classes of constraints emerge: (i) Only propagating modes of resolution $\Delta p\le|\phi|$ need to be considered. 
(ii) Since coarse-graining generates (quasi)particle 
masses for $\Delta p\le |\phi|$ we need assure that 
the exchange of unresolved massless particles contributing to an 
effective, local vertex does not involve momentum 
transfers larger than $|\phi|$. Condition (i) reads
%*****
\eqb
\label{cond1}
|p^2-m^2|\le |\phi|^2\,\ \ \ (\mbox{massive mode})\,,\ \ \ \ \ 
|p^2|\le |\phi|^2\,\ \ \ (\mbox{massless mode})\,
\eqe
%******
where $|\phi|=\sqrt{\frac{\Lambda^3}{2\pi T}}$. 
For a three-vertex (ii) is contained in (i) 
by momentum conservation. For a four-vertex condition (ii) distinguishes 
$s$, $t$, and $u$ channels in the scattering process. Labelling the ingoing (outgoing) momenta 
by $p_1$ and $p_2$ ($p_3$ and $p_4=p_1+p_2-p_3$), we have
%**********
\eqb
\label{cond2}
|(p_1+p_2)^2|\le|\phi|^2\,,\ \ (s)\ \ |(p_3-p_1)^2|\le|\phi|^2\ \ (t)\,,\ \ |(p_2-p_3)^2|\le|\phi|^2\ \ (u)\,.
\eqe
%**********
Notice conditions (\ref{cond2}) reduce to 
the first condition if one computes the one-loop 
tadpole contribution to the polarization tensor or the four-vertex induced 
two-loop contribution to a thermodynamical quantity\footnote{The $t$-channel condition is then trivially satisfied 
while the $u$-channel condition reduces to the $s$-channel condition by letting 
the loop momentum $k\to -k$ in $|(p-k)^2|\le |\phi|^2$, 
see \cite{Hofmann2005,HerbstHofmannRohrer2004,SchwarzHofmannGiacosa20061}.}. The pressure was computed up to two loops in 
\cite{Hofmann2005,HerbstHofmannRohrer2004,SchwarzHofmannGiacosa20061}: Two-loop corrections are smaller than 
(depending on temperature) $\sim 0.1\%$ of the one-loop result.

\begin{figure}
\begin{center}
\leavevmode
\leavevmode
%\epsffile[80 25 534 344]{}
\vspace{1.6cm}
\includegraphics{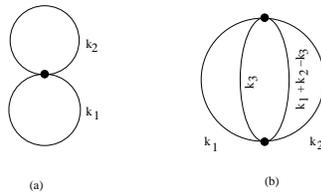}
\end{center}
\caption{\protect{\label{Fig-2}} (a) Two-loop and (b) three-loop diagram contributing 
to the pressure in the deconfining phase of SU(2) Yang-Mills thermodynamics. 
The solid lines refer to thermal massive-mode propagation.}      
\end{figure}

We expect that the contribution of $N$-particle irreducible ($N$PI) polarizations to the dressing of propagators 
vanishes for $N>N_{\tiny\mbox{max}}<\infty$ since (\ref{cond1}) and (\ref{cond2}) then 
impose more independent conditions than there are independent 
loop-momentum components. It is instructive to analyze the two bubble diagrams 
in Fig.\,\ref{Fig-2}. While, due to (\ref{cond1}) and (\ref{cond2}), the two-dimensional 
region of integration for $|\vec{k_1}|$ and $|\vec{k_2}|$ in diagram (a) is non-compact the three-dimensional 
region of integration for $|\vec{k_1}|$, $|\vec{k_2}|$, and $|\vec{k_3}|$ is compact 
in diagram (b) \cite{Hofmann20062}.    
%***********************
%************************   
In one-particle reducible diagrams 
so-called pinch-singularities arise in the 
real-time dressing of propagators (powers of delta functions). But a re-summation 
of 1PI polarizations modifies
\footnote{To avoid a logical contradiction the 1PI polarizations are first computed in 
real time subject to the constraints (\ref{cond1}) and 
(\ref{cond2}). Subsequently, a continuation to imaginary time in the external momentum variable $p^0$ 
is performed. Then the re-summation is carried out, and finally the result is 
continued back to real time.} the scalar part of the tree-level propagators by 
momentum dependent screening functions with finite imaginary parts. 
This makes powers of spectral functions well-defined. To summarize, 
the effective loop expansion should be given by 
infinite re-summations of a finite number of $N$PI polarizations \cite{Hofmann20062}. 
Because the latter dramatically decrease with $N$ 
radiative effects are reliably approximated at the two-loop level.

\section{Application: SU(2)$_{\tiny\mbox{CMB}}$}\label{sec4}

A (falsifiable, see below) postulate emerges\cite{SchwarzHofmannGiacosa20061,Hofmann2005,Hofmann20051,GiacosaHofmann2005}: 
SU(2) Yang-Mills dynamics of scale $T_{\tiny\mbox{CMB}}\sim \Lambda_{\tiny\mbox{CMB}}=1.065\times 10^{-4}\,$eV 
(thus the name SU(2)$_{\tiny\mbox{CMB}}$) masquerades today as the U(1)$_Y$ of the Standard Model (SM). 
$\Lambda_{\tiny\mbox{CMB}}$ derives from the 
b.c. that light propagates in an 
unadulterated way today\footnote{This is the case only for 
$T_{\tiny\mbox{CMB}}=T_c$.}. All low-temperature 
($T\ll 0.5\,$MeV) dynamics of the SM with momentum transfers considerably 
below the `electroweak scale' $\sim 200\,$GeV is unaffected by this 
assignment if one distinguishes between propagating (the massless excitations of SU(2)$_{\tiny\mbox{CMB}}$) and interacting 
photons\footnote{Interaction with electroweak matter 
dynamically invokes the Weinberg angle by a rotation 
of the propagating to the interacting photon, for a discussion 
see \cite{GiacosaHofmann2005}.}. An exception takes place for $T$ a few times 
$T_{\tiny\mbox{CMB}}=2.351\times 10^{-4}\,$eV$\sim 2.73\,$K
\footnote{Due to interactions with the massive excitations of SU(2)$_{\tiny\mbox{CMB}}$ referred to as $V^\pm$ in the following.}. 
Namely, the photon's dispersion law modifies, see \cite{SchwarzHofmannGiacosa20061}:
%*********
\eqb
\label{displaw}
\omega^2(\vec{p})=\vec{p}^2\ \ \longrightarrow \ \ \omega^2(\vec{p},T)=\vec{p}^2+G(\omega(\vec{p},T),\vec{p},T)\,.
\eqe
%*********
At photon momenta $p>0.2...0.3\,T$ small antiscreening takes place ($G<0$ in Eq.\,(\ref{displaw})) which  
dies off exponentially with $p$. There 
is a power suppression of $|G|$ with increasing $T$. For $p$ being a small fraction of $T$ ($T\sim 5$\,K $\Rightarrow$ $p\le\,0.2\,T$) 
antiscreening converts into screening 
($G>0$) which rapidly grows for decreasing $p$. The effect is negligible 
for photon-gas temperatures sufficiently above $T_{\tiny\mbox{CMB}}$, 
say $T>80\,K$, and it is absent at $T=T_{\tiny\mbox{CMB}}$, see 
\cite{SchwarzHofmannGiacosa20061}. A modification of the black-body spectrum emerges 
at low temperatures and low momenta \cite{SchwarzHofmannGiacosa20062}: 
At $T=10\,$K the spectral intensity vanishes for frequencies $0<\omega\le 0.12\,T$. For $0.12\,T\le\omega\le 0.25\,T$ there is 
excess of spectral power. This prediction tests the postulate 
SU(2)$_{\tiny\mbox{CMB}}\stackrel{\tiny\mbox{today}}=$U(1)$_Y$. 
The relative deviation to the U(1) black-body pressure peaks at $T\sim 2\,T_{\tiny\mbox{CMB}}$ on the 
$10^{-3}$-level coinciding with the strength 
of the CMB dipole\footnote{We expect that besides the contribution due to the Doppler-effect \cite{Wilkinson} also 
a dynamical part to generate the CMB dipole.}. The observation of cold, {\sl old}, and dilute 
clouds of atomic hydrogen in between the spiral arms of our galaxy \cite{BruntKnee} hints 
to SU(2)$_{\tiny\mbox{CMB}}\stackrel{\tiny\mbox{today}}=$U(1)$_Y$ being true 
\cite{SchwarzHofmannGiacosa20061,SchwarzHofmannGiacosa20062}\footnote{The forbidden wavelengths 
at brightness temperature $T=5\,$K \cite{BruntKnee}, range from 2.1\,cm to 
8.8\,cm. This is comparable to the mean 
distance between H-atoms \cite{BruntKnee}: The dipole force, which 
would cause all atoms to convert into H$_2$ molecules within a time two orders 
of magnitude lower than the 
inferred age of the cloud, is switched off.}. The suppression of 
low-momentum photons could be the reason for the missing power 
in TT CMB spectra at low $l$ \cite{DSchwarz2006}\footnote{The suppression 
of `messenger' photons weakens the 
correlation between temperature fluctuations at large angular separation in the sky.}. Also there 
is the result of the PVLAS experiment \cite{PVLAS}\footnote{A dichroism induced by a 5\,Tesla 
homogeneous magnet on linearly polarized laser light with the temperature of the apparatus being $\sim 4.2$\,K. 
When fitted to an axion model the inferred 
axion mass is about 1\,meV with a too large coupling 
(contradicting solar bounds on axion-induced $X$-ray emission \cite{Masso2006}). The observation making contact 
with SU(2)$_{\tiny\mbox{CMB}}$ is that at $T=4.2\,$K one has $m_{V^\pm}=0.4\,$meV: 
A value comparable to the axion mass. It is the mass of the 
propagating intermediary particle which is a nearly model independent 
quantity in non-standard theories of photon-photon coupling. Thus $m_{V^\pm}=0.4\,$meV is an 
encouraging observation. Moreover, the small photon-to-photon 
coupling would be explained by the smallness of a kinematically strongly 
constrained loop-propagation of $V^\pm$ excitations, see Sec.\,\ref{sec2}.} 
To quantitatively investigate this the two one-loop diagrams for the photon polarization involving the full 
$V^\pm$-propagator in the external magnetic field must be calculated. 
We hope to tackle this task in the near future. 

\bibliographystyle{ws-procs9x6}
\bibliography{ws-pro-sample}

\end{document}